\documentclass[figures]{epl}
\usepackage{amsmath}
\usepackage{amssymb}

\title{Synchronization and frustration in oscillator
networks with attractive and repulsive interactions}
\shorttitle{Synchronization and frustration in oscillator
networks}
\author{Dami\'an H. Zanette}
\shortauthor{D. H. Zanette} \institute{Consejo Nacional de
Investigaciones Cient\'{\i}ficas y T\'ecnicas, Centro At\'omico
Bariloche and Instituto Balseiro, 8400 Bariloche, R\'{\i}o Negro,
Argentina }


\begin{document}

\maketitle

\begin{abstract}
We study, numerically and analytically, the stability of
synchronization for an ensemble of coupled phase oscillators with
attractive and repulsive interactions, as a function of the number
of repulsive couplings and their intensity. Scaling properties of
the desynchronization transition are disclosed, and it is shown
that unsynchronized configurations have different symmetries
depending on the intensity of the repulsive interaction. The
concept of frustration minimization helps understanding the main
features observed in the collective dynamics of the oscillator
network.
\end{abstract}

Synchronization is a well-known form of collective dynamics in
large ensembles of interacting dynamical systems. Under the action
of attractive coupling, strong correlations develop in time
between the states of individual elements, and the ensemble is
able to generate signals of macroscopic amplitude. This  coherent
behaviour is found in a broad class of natural systems
--especially, in the realm of life sciences \cite{Win}-- and has
been reproduced by means of a variety of mathematical models
\cite{Kura,pik,nos}.

The strongest manifestation of synchronized dynamics, which can be
realized in an ensemble of identical periodic oscillators subject
to attractive coupling, is full synchronization. In a fully
synchronized ensemble, the individual motions of all the
oscillators coincide. Heterogeneities, chaotic individual
dynamics, and repulsive interaction, on the other hand, may
inhibit the development of synchronization \cite{nos}.

In this Letter, we explore the competing effect of attractive and
repulsive interaction in oscillator ensembles. Specifically, we
study the influence of repulsive couplings on the stability of
full synchronization. The combined action of attractive and
repulsive couplings has already been considered by Daido
\cite{d1,d2,d3}, who disclosed the possibility of weaker forms of
synchronized dynamics, glass-like behaviour, and anomalous
relaxation. These results, however, typically apply to cases where
the number of attractive and repulsive couplings and their
relative intensity are equal on the average. Here, instead, we are
interested at the effect of repulsive couplings as their abundance
and intensity are varied gradually. As discussed below, this leads
naturally to the consideration of heterogeneous random networks of
coupled oscillators.

We consider an ensemble of $N$ identical oscillators, whose
individual states are given by the phases $\phi_i \in [0,2\pi)$
($i=1,\dots,N$). In the absence of interaction, the evolution of
the phases is given by $\dot \phi=\omega$ for all $i$. Coupling is
introduced following Kuramoto's scheme \cite{Kura},
\begin{equation} \label{osc}
\dot \phi_i = \omega + \frac {1}{N} \sum_{j=1}^N W_{ij} \sin
(\phi_j-\phi_i),
\end{equation}
where $W_{ij}$ weights the interaction between oscillators $i$ and
$j$. By transforming $\phi_i \to \phi_i + \omega t$  for all $i$,
we fix, without generality loss, $\omega=0$. In the following, we
study the case of symmetric coupling, $W_{ij}=W_{ji}$. For each
realization of the ensemble, the interaction weights are chosen
at random as
\begin{equation}
W_{ij} = \left\{
\begin{array}{rl}
1 & \mbox{with probability $1-p$} \\ -w & \mbox{with probability
$p$}
\end{array}
\right.
\end{equation}
with $w>0$. Thus, on the average, a fraction $p$ of the $N(N-1)/2$
couplings correspond to repulsive interaction, while the remaining
couplings are attractive. The relative intensity of the repulsive
interaction is $w$. Equations (\ref{osc}) can be rewritten as
\begin{equation} \label{osc1}
\dot \phi_i=\frac{1}{N} \sum_{j=1}^N \sin (\phi_j-\phi_i) -
\frac{1+w}{N}\sum_{j\in {\cal R}_i} \sin (\phi_j-\phi_i).
\end{equation}
In this representation, the first term in the r.h.s.~of the
equation corresponds to global attractive coupling, where all
oscillator pairs interact with the same intensity. The second term
corresponds to repulsive interaction of relative intensity $1+w$.
Repulsive couplings are restricted to a random network determined
by the sets ${\cal R}_i$. An oscillator $j$ belongs to ${\cal
R}_i$ if it interacts repulsively with oscillator $i$. Calling
$z_i$ the number of oscillators in ${\cal R}_i$, the average
connectivity of the network of repulsive couplings is $\langle z_i
\rangle = p(N-1)$.

Our aim here is to analyze the long-time solutions of Eqs.
(\ref{osc1}) as the parameters $p$ and $w$ are varied, averaging
over different realizations of the network of repulsive couplings.
Specifically, we are interested at the destabilization of full
synchronization as the number and intensity  of repulsive
couplings grow. The state of full synchronization,
$\phi_i(t)=\phi^*$ for all $i$ and constant $\phi^*$, is in fact a
solution to Eqs. (\ref{osc1}) for any $w$ and arbitrary ${\cal
R}_i$. Its stability, however, is expected to depend on the
intensity of the repulsive interaction and on the network
topology. For $p=0$, which corresponds to the case of pure global
attractive coupling, repulsive interaction is absent and full
synchronization is stable. For $p=1$, coupling is also global but
purely repulsive; the sets ${\cal R}_i$ are extended to the whole
ensemble. In this situation, full synchronization is unstable and,
for asymptotically large times, the system reaches one of the
infinitely many states given by the equations $\sum_j \sin
(\phi_j-\phi_i)=0$ ($i=1,\dots,N$). In such states, typically, the
phases $\phi_i$ are more or less uniformly distributed over
$[0,2\pi)$ \cite{nos}. Destabilization of the fully synchronized
state is thus expected to occur at intermediate values of $p$.

Linear stability of full synchronization can be assessed by
studying the eigenvalue spectrum of the $N\times N$ matrix with
elements
\begin{equation} \label{Sij}
S_{ij}=\frac{1}{N}+ \left( -1+\frac{1+w}{N} z_i \right)
\delta_{ij} - \frac{1+w}{N} R_{ij},
\end{equation}
where $\delta_{ij}$ are the elements of the identity matrix, and
$R_{ij}$ are the elements of the (symmetric) adjacency matrix
corresponding to the network of repulsive couplings: $R_{ij}=1$ if
$j\in {\cal R}_i$, and $0$ otherwise. The state of full
synchronization is linearly stable if {\it all} the eigenvalues of
$S_{ij}$ are negative.

\begin{figure}
\onefigure[width=8cm]{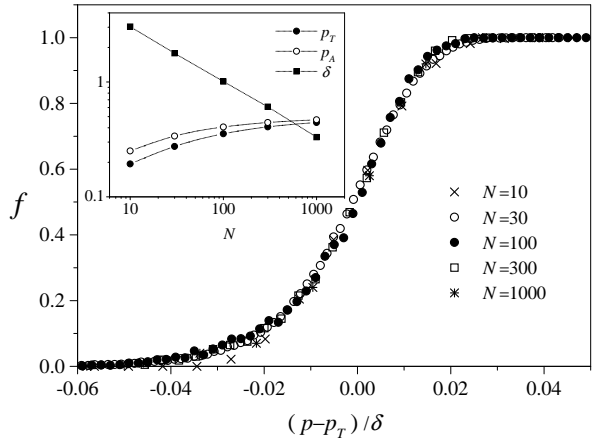} \caption{Data collapse of the
fraction $f$ of realizations with unstable full synchronization,
as a function of the fraction of repulsive couplings $p$, for
different system sizes $N$. The intensity of the repulsive
interaction is $w=1$. Inset: the transition point $p_T$ and the
transition width $\delta$ as  functions of $N$. The analytical
result for the transition point, $p_A$, is also shown. Dotted
lines are spline interpolations, added for clarity.} \label{fig1}
\end{figure}

Using a standard numerical routine, we have calculated the
eigenvalues of $S_{ij}$ for different values of the parameters $p$
and $w$, and system sizes $N$ between $10$ and $10^3$. For each
parameter set, $10^2$ to $10^4$ realizations of the network of
repulsive couplings have been considered. As a measure of the
degree of instability of full synchronization, we have recorded
the fraction $f$ of realizations for which at least one of the
eigenvalues is positive --i.e., for which full synchronization is
unstable. For fixed $N$ and $w$, as expected, this fraction
changes from $f\approx 0$ for small $p$ to $f\approx 1$ for large
$p$. The transition takes place in a rather narrow range of $p$.
As illustrated in Fig. \ref{fig1} for $w=1$, numerical results for
$f$ as a function of $p$ and different system sizes admit a fairly
good collapse as a function of the transformed parameter
$(p-p_T)/\delta$. Here, $p_T$ has been chosen as the value of $p$
for which $f=1/2$, which we identify as the transition point. The
coefficient $\delta$, in turn, measures the width of the
transition range. For each size $N$, this coefficient is adjusted
to achieve the data collapse, taking as a reference --without
generality loss-- $\delta=1$ for $N=100$. The inset of Fig.
\ref{fig1} shows a plot of $p_T$ and $\delta$ as functions of $N$.
As $N$ grows, the transition point seems to approach an asymptotic
value $p_T\approx 0.5$. The width $\delta$, on the other hand,
decreases steadily, approximately following a power law
$N^{-0.48\pm 0.01}$. This indicates that the transition range
becomes narrower, and the transition itself becomes better
defined, as the system size is increased.

An independent, analytical evaluation of the transition point can
be achieved under the hypothesis that the network of repulsive
couplings is regular, i.e. when all the sets ${\cal R}_i$ have
exactly the same number of elements, $z_i=z=p(N-1)$ for all $i$.
In these conditions,  the eigenvalues of $S_{ij}$ in Eq.
(\ref{Sij}) are directly related to the eigenvalues $\rho_k$
($k=1,\dots,N$) of $R_{ij}$. Consequently, the stability condition
for full synchronization can be given in terms of $\rho_k$. The
transition point turns out to be determined by the minimum
eigenvalue of $R_{ij}$, $\rho_{\min}= \min \{ \rho_k\}$. For
sufficiently large $N$, the typical value of this eigenvalue can
be estimated from the so-called semicircle law \cite{semic} which,
in the present framework, establishes that the spectral density of
$R_{ij}$ has the shape of a semicircle of radius $2\sqrt{Np(1-p)}$
centered at the origin. The instability threshold for full
synchronization is given by the condition \cite{fut}
\begin{equation} \label{anal}
p\left( 1-\frac{1}{N}\right)+ 2 \sqrt{\frac{p(1-p)}{N}}
=\frac{1}{1+w}.
\end{equation}
The inset of Fig. \ref{fig1} shows that the value $p_A$ obtained
from this equation for $w=1$, as a function of the system size, is
in very good qualitative agreement with $p_T$. Note that,  for $N
\to \infty$, our analytical evaluation predicts a transition point
at $p_A=(1+w)^{-1}$; for $w=1$, we get $p_A=1/2$. Evaluation of
the spectral density of $R_{ij}$ beyond the semicircle law
\cite{Bronk, Farkas} shows that the fraction of eigenvalues
outside the interval $(-2\sqrt{Np(1-p)},2\sqrt{Np(1-p)})$
decreases with the system size as $N^{-1}$. This is consistent
with our observation that the transition becomes sharper as $N$
grows.

\begin{figure}
\onefigure[width=8cm]{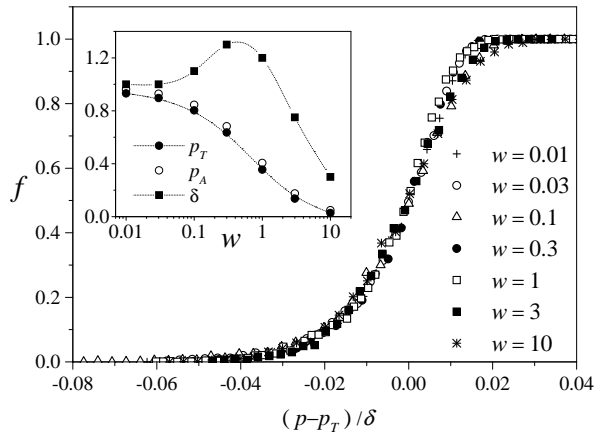} \caption{Data collapse of the
fraction $f$ of realizations with unstable full synchronization,
as a function of the fraction of repulsive couplings $p$, for
different intensities   $w$. The system size is $N=100$. Inset:
the numerical and the analytical transition points, $p_T$ and
$p_A$, and the transition width $\delta$, as functions of $w$.
Dotted lines are spline interpolations, added for clarity.}
\label{fig2}
\end{figure}

Figure \ref{fig2} shows numerical results for  $f$ in systems of
fixed  size, $N=100$, for various values of $w$. As in Fig.
\ref{fig1}, the data collapse was achieved by plotting $f$ as a
function of $(p-p_T)/\delta$, where $p_T$ is the fraction of
repulsive couplings at which $f=1/2$. The inset shows $p_T$ and
$\delta$ as functions of $w$. Since for both large and small $w$
the transition range is narrower than in between, $\delta$ reaches
a maximum for an intermediate value of the intensity of the
repulsive interaction. The analytical threshold $p_A$, obtained
from Eq. (\ref{anal}), is again in very good qualitative agreement
with the numerical results.

\begin{figure}
\onefigure[width=8cm]{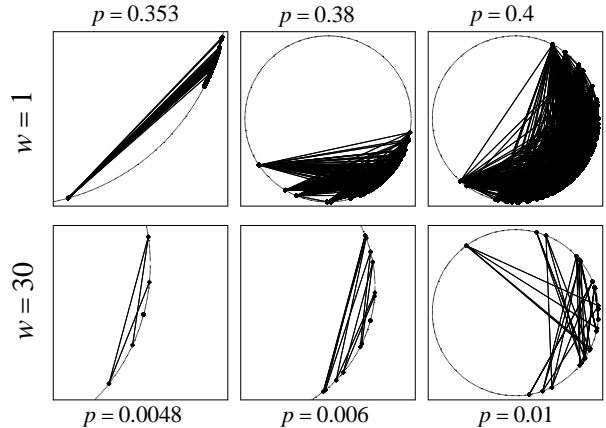} \caption{Long-time snapshots of
the distribution of oscillator phases for different values of the
intensity $w$ and fraction $p$ of repulsive couplings, in a system
of size $N=100$. Individual phases are represented by dots on the
unit circle (dotted line). Straight lines join oscillators which
interact repulsively.} \label{fig3}
\end{figure}

What is the nature of the stationary state of the oscillator
ensemble just beyond the destabilization of full synchronization?
To answer this question, we have inspected the long-time
distribution of phases for networks of $N=100$ oscillators, where
repulsive couplings were successively added. This has made it
possible to detect, for each network, the exact point at which
full synchronization becomes unstable, and to analyze the
resulting unsynchronized asymptotic state. It turns out that this
unsynchronized state changes qualitatively when the intensity of
the repulsive interaction grows. For small and moderate $w$,
destabilization of full synchronization gives place to a state
where oscillator phases are irregularly spread over a small
interval. Typically,  the oscillator with maximal number of
repulsive couplings has a substantially different phase. As
expected, the distribution of phases widens as the fraction of
repulsive couplings  $p$ increases. These features are illustrated
by the upper panels of Fig. \ref{fig3}, for $w=1$. In this
particular realization of the oscillator network, full
synchronization becomes unstable for $p=0.3527\dots$. The figure
shows snapshots of the phase distribution in the plane $(\cos
\phi, \sin \phi)$. Lines join oscillator pairs with repulsive
couplings.

The situation changes substantially if the intensity of the
repulsive interaction grows enough. For large $w$, when relatively
few repulsive couplings are needed to destabilize full
synchronization, unsynchronized states just beyond the transition
are characterized by a symmetric distribution of phases. Some of
the  oscillator pairs with repulsive interaction abandon the
synchronized cluster and their two elements reach symmetric
positions at each side of the cluster. The remaining oscillators
stay synchronized at a common phase. As the fraction of repulsive
couplings keeps growing, the number of unsynchronized pairs
increases, still preserving their symmetric configuration.
Eventually, however, symmetry is broken, and the ensemble reaches
a more or less irregular phase distribution, similar to that
observed for lower $w$. The second line of panels in Fig.
\ref{fig3} illustrates these features for $w=30$. In this
realization of the network, the transition takes place at
$p=0.0048$, corresponding to just $24$ repulsive couplings. For
that value of $p$ and for $p=0.006$ we find symmetric
configurations of phases, while for $p=0.01$ the distribution is
already asymmetric. In the symmetric cases, the cluster of
remaining synchronized oscillators is the isolated dot in the
centre of the configuration.

Our numerical findings on the destabilization of full
synchronization and the resulting  unsynchronized states can be
understood, at least qualitatively, by resorting to a concept
borrowed from the study of disordered spin systems, namely, {\it
frustration} \cite{frustration,d2}. In spin systems, where
interactions are usually local, frustration is associated with the
impossibility that three mutually interacting spins with
interaction weights of different sign achieve simultaneously a
state of minimal energy. In our system, where (attractive or
repulsive) interaction affects every oscillator pair, frustration
must be defined for each pair in relation to the state of the
whole ensemble. For instance, an oscillator pair with repulsive
coupling is frustrated if the system is in a collective
synchronized state, so that the individual phases of the two
oscillators are forced to coincide. Conversely, a pair with
attractive coupling is frustrated if the interactions with the
rest of the ensemble inhibits the synchronization of the two
oscillators. Frustration is quantitatively characterized by the
function
\begin{equation} \label{Frust}
F=-\frac{1}{N}\sum_{i,j=1}^N W_{ij} \cos (\phi_j-\phi_i).
\end{equation}
Pairs with attractive coupling, $W_{ij}>0$, contribute to decrease
the frustration $F$ if $\phi_j\approx \phi_i$. {\it Vice versa},
for $W_{ij}<0$, $F$ decreases as the difference of the two phases
grows. The relevance of the frustration $F$ with respect to the
oscillator dynamics resides in the fact that it plays the role of
a nonequilibrium potential. Specifically, for $\omega=0$, Eqs.
(\ref{osc}) can be written as $\dot \phi_i =-\partial F/\partial
\phi_i$. Therefore,
\begin{equation}
\dot F = \sum_{i=1}^n \frac{\partial F}{\partial \phi_i} \dot
\phi_i=  -\sum_{i=1}^n \left( \frac{\partial F}{\partial \phi_i}
\right)^2 \le 0 .
\end{equation}
In other words, the dynamics convey a steady decrease of the
frustration.

Bearing these considerations in mind, we can argue as follows. As
the fraction of repulsive couplings grows from $p=0$, full
synchronization is initially stable. Frustration, nevertheless,
increases because of the increasing number of oscillators with
repulsive interactions that are forced to stay into the
synchronized cluster. The collective state of the ensemble is thus
able to bear a certain degree of frustration, in order to maintain
synchronization. Eventually, however, frustration reaches too high
levels, and synchronization is no longer a ``convenient''
collective configuration --though it still is a possible
stationary state for the ensemble. Consequently, the synchronized
cluster breaks down, the individual oscillator phases spread out,
and the level of frustration is alleviated. Since, now,
oscillators are not locked to the synchronized cluster, further
increase of $p$ will lead to the gradual widening of the
distribution of phases, in the search for the state of minimal
frustration.

When the repulsive interaction is strong enough, just a few
repulsive couplings suffice to make full synchronization
``inconvenient'' from the viewpoint of frustration, and the
synchronized cluster may break down for very small values of $p$.
Under these conditions, the network of repulsive couplings is very
simple: it typically consists of isolated oscillator pairs. If
full synchronization is unstable, the two oscillators of each pair
abandon the synchronized cluster, trying to maximize their mutual
distance. At the same time, both of them tend to minimize the
distance to the cluster, since their interaction with the
oscillators there is attractive. The resulting configuration is
thus symmetric around the cluster, as illustrated in Fig.
\ref{fig3} for $w=30$.

In summary, we have studied the desynchronization transition in a
network of coupled identical phase oscillators with attractive and
repulsive interactions, as the number and intensity of repulsive
couplings is increased. For finite system sizes, the transition is
gradual: the probability that the state of full synchronization is
unstable for a given realization of the network of couplings grows
smoothly with the number of repulsive couplings. Numerical
finite-size analysis as well as analytical results, however,
suggest that the desynchronization transition would become abrupt
for infinitely large systems. Beyond the transition, the nature of
the unsynchronized long-time state depends sensibly on the
intensity of the repulsive interaction. For small and moderate
intensities, individual oscillator phases reach a more or less
disordered configuration, reminiscent of the disordered
equilibrium state of an oscillator ensemble with global repulsive
coupling \cite{nos}. On the other hand, for strong repulsive
interaction, desynchronization gives rise to highly symmetric
configurations, where the phase of unsynchronized oscillators are
situated at both sides of the remaining synchronized cluster. The
possibility of observing these symmetric configuration depends on
their existence and stability, which we analyze in detail
elsewhere \cite{fut}. The notion of frustration --already
addressed, in the frame of coupled oscillator ensembles, by Daido
\cite{d2}-- is useful to construct a variational-like formulation
for the dynamics of the system studied here. In this framework, we
have given a qualitative description of our main results. Such
formulation, however, should also play a role in a quantitatively
detailed analysis of the system. In fact, as defined in Eq.
(\ref{Frust}) and under appropriate conditions at its singular
points, frustration could be used as a Lyapunov function for the
dynamics.

\end{document}